\documentclass[aip,amsmath,amssymb,reprint]{revtex4-1}

\usepackage{graphicx}
\usepackage{subfig}
\usepackage{dcolumn}
\usepackage{comment}
\usepackage{bm}
\usepackage[utf8]{inputenc}
\usepackage[T1]{fontenc}
\usepackage{mathptmx}
\usepackage[colorlinks=true,linkcolor=blue,citecolor=blue,urlcolor=blue]{hyperref}

\begin{document}

\preprint{AIP/123-QED}

\title[Superconducting Nanowire Single Photon Detectors based on NbRe nitride ultrafilms]{Superconducting Nanowire Single Photon Detectors based on NbRe nitride ultrathin films}

\author{F. Avitabile}
\affiliation{CNR-SPIN, c/o Università degli Studi di Salerno, I-84084 Fisciano (Sa), Italy}
\author{F. Colangelo}
\affiliation{Dipartimento di Fisica ``E.R. Caianiello'', Università degli Studi di Salerno, I-84084 Fisciano (Sa), Italy}
\affiliation{CNR-SPIN, c/o Università degli Studi di Salerno, I-84084 Fisciano (Sa), Italy}
\affiliation{Department of Imaging Physics (ImPhys), Faculty of Applied Sciences, Delft University of Technology, Delft 2628 CJ, The Netherlands}
\author{M. Yu. Mikhailov}
\affiliation{Department of Imaging Physics (ImPhys), Faculty of Applied Sciences, Delft University of Technology, Delft 2628 CJ, The Netherlands}
\author{Z. Makhdoumi Kakhaki}
\affiliation{Dipartimento di Fisica ``E.R. Caianiello'', Università degli Studi di Salerno, I-84084 Fisciano (Sa), Italy}
\affiliation{CNR-SPIN, c/o Università degli Studi di Salerno, I-84084 Fisciano (Sa), Italy}
\author{A. Kumar}
\affiliation{Dipartimento di Fisica ``E.R. Caianiello'', Università degli Studi di Salerno, I-84084 Fisciano (Sa), Italy}
\affiliation{CNR-SPIN, c/o Università degli Studi di Salerno, I-84084 Fisciano (Sa), Italy}
\author{I. Esmaeil Zadeh}
\affiliation{Department of Imaging Physics (ImPhys), Faculty of Applied Sciences, Delft University of Technology, Delft 2628 CJ, The Netherlands}
\author{C. Attanasio}
\affiliation{Dipartimento di Fisica ``E.R. Caianiello'', Università degli Studi di Salerno, I-84084 Fisciano (Sa), Italy}
\affiliation{CNR-SPIN, c/o Università degli Studi di Salerno, I-84084 Fisciano (Sa), Italy}
\affiliation{Centro NANO\_MATES, Università degli Studi di Salerno, I-84084 Fisciano (Sa), Italy}
\author{C. Cirillo}
\thanks{Corresponding author carla.cirillo@spin.cnr.it}
\affiliation{CNR-SPIN, c/o Università degli Studi di Salerno, I-84084 Fisciano (Sa), Italy}

\date{\today}

\begin{abstract}
The influence of the reactive DC sputtering parameters on the superconducting properties of NbReN  ultrathin films was investigated. A detailed study of the current-voltage characteristics of the plasma was performed to optimize the superconducting critical temperature, $T_{\mathrm{c}}$. The thickness dependence of $T_{\mathrm{c}}$ for the films deposited under different conditions was analyzed down to the ultrathin limit. Optimized films were used to fabricate superconducting nanowire single photon detectors which, at $T=3.5$~K, show saturated internal detection efficiency (IDE) up to a wavelength of 1301 nm and 95\% IDE at 1548 nm  with recovery times and timing jitter of about 8~ns and 28~ps, respectively.
\end{abstract}

\maketitle


Superconducting nanowire single photon detectors (SNSPDs)~\cite{ZadehRev} are one of the key enabling technologies of many quantum-based application, from computation to cryptography~\cite{You}. Other strategic operation fields include space to ground communication~\cite{Hao2024}, monitoring of atmospheric pollution~\cite{Salvoni2022}, medical sensing~\cite{Ozana,Tamami}, while further unconventional applications are still at their earlier stage~\cite{Polakovic}. As a consequence, many efforts are put to enhance specific performance of these devices, depending on their operation field~\cite{Holzman2019}. New platforms and configurations, for example superconductor/superconductor bilayers~\cite{Ivry2017}, Josephson junctions-based single photon detectors~\cite{Walsh2021}, even integrated with innovative materials such as graphene~\cite{Walsh2017}, or carbon nanotubes~\cite{Rampini2024} were proposed. At the same time, material-oriented research makes continuous progress in the development of these devices~\cite{Holzman2019,Lita2015}. For example, in order to meet specific application requirements \cite{Cheng2020,Stepanov2024,Zichi2019,Shan2021,Chang2022}, research still focuses on the optimization of crystalline superconductors with high critical temperature, $T_{\rm c}$, such as NbN and NbTiN ($T_{\rm c} \approx 16$ K), despite their excellent performance as SNSPDs at 1550 nm. Therefore, ion bombardment during sputtering deposition was used to obtain NbN films with increased low temperature resistivity ($\rho$) and reduced $T_{\rm c}$ for the realization of SNSPDs with 100\% detection efficiency typical of devices made of amorphous superconductors, with the excellent timing performance and higher operating temperatures characteristic of polycrystalline detectors~\cite{Dane2017}. Moreover, alternative superconductors are constantly suggested. Indeed, recent studies have explored crystalline nitrides such as MoN~\cite{Hallett2021} and VN~\cite{Zolotov2021}. Similarly, we are currently investigating NbReN, a new nitride superconductor recently synthesized in form of polycrystalline films with grains of small dimensions, typically 2–3 nm, and bulk $T_{\rm c}$ of about 5 K~\cite{Cirillo2021}. Its electrical transport properties, typical of dirty superconductors, may in principle be of interest in the field of superconducting electronics, in particular for the realization of SNSPDs or high kinetic inductance devices~\cite{Moshe}. Its parent compound NbRe~\cite{Cirillo2016,Caputo2017} already demonstrated high performances for the realization of single photon detectors both in forms of nanometric meanders~\cite{Cirillo2020} and microstrips~\cite{Ejrnaes2022,Ercolano2023,Cirillo2024}. Compared with NbRe, NbReN films feature higher resistivity values, and reduced superconducting critical temperature~\cite{Cirillo2021}, comparable to those of amorphous superconductors~\cite{Verma2021,Colangelo2022,Chen2021}. These characteristics may reasonably offer the opportunity to operate NbReN-based detectors in the mid-infrared (MIR) spectral range. In the MIR regime NbReN could be competitive with amorphous superconductors, such as WSi~\cite{Verma2021,Colangelo2022} or MoSi~\cite{Chen2021}, which typically suffer from poor time resolution and very low operation temperature. In contrast, NbReN-based detectors are expected to present excellent time performances, as its parent compound~\cite{Cirillo2021}, due to the low values of the quasiparticle relaxation times estimated from flux-flow instability study ~\cite{Kakhaki2024}. Here, we report on the first SNSPD based on NbReN films. First we illustrate the simple yet reliable deposition process developed to correlate the sputtering conditions to the electrical properties of the films both in the normal and superconducting state. The deposition conditions were tuned to produce ultrathin NbReN films with $T_{\rm c}$ values suitable to operate the detector at temperatures reachable with commercial cryogen free systems. As a result, nanometric meanders, with wires about 70 nm wide and circular detection area with a diameter of about 4 $\mu$m based on 10-nm-thick NbReN films, show 95\% IDE at 1550 nm and excellent time performance at 3.5 K. 

\begin{table}
\noindent \begin{centering}
\begin{tabular}{|c|c|c|c|c|}
\hline 
series name  & $P_{\mathrm{tot}}$($\mu$bar) & N$_{2}^{\%}$($\mu$bar)  & P(W)  & $d_{\mathrm{NbReN}}$(nm) \tabularnewline
\hline 
\hline 
A$^{\mathrm{ Ref.}}$~\cite{Cirillo2021}& $4.5-13$ & $23-44$  & $100-150$  & $4-70$\tabularnewline
\hline 
B & $3.0-4.2$ & $9-29$  & $150-350$  & $5-50$\tabularnewline
\hline 
C & $3.5-6.3$ & $2.8-6.0$ & $40-200$  & $6-50$\tabularnewline

\hline  
\end{tabular}
\par\end{centering}
\caption{Deposition conditions and film thicknesses for the series of samples under investigation\label{tab:samples}}
\end{table}

\begin{figure*}[!ht]
\centerline{\includegraphics[width=18.5cm]{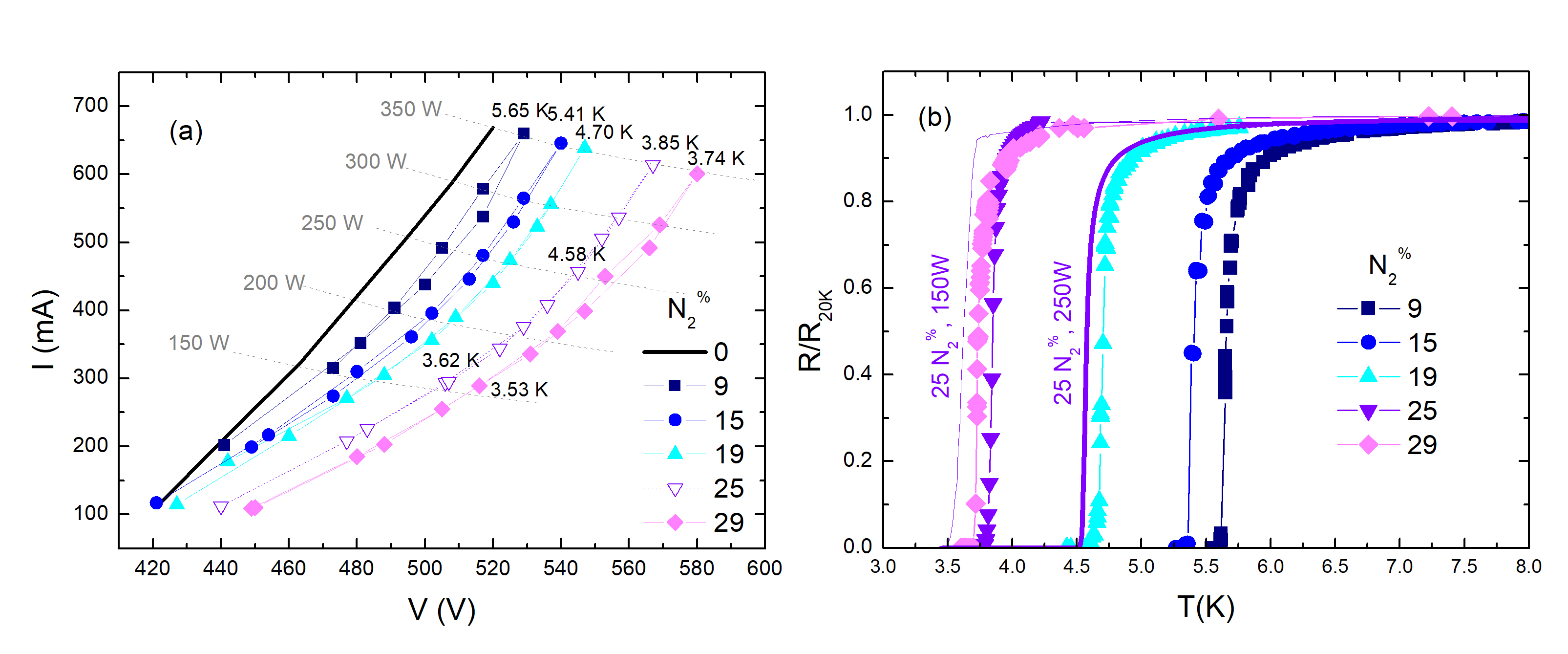}}
\caption{(a) Current-Voltage characteristics for $p_{\mathrm{Ar}}=3$~$\mu$bar and different N$_{2}^{\%}$ by varying the sputtering power, $P$. The black curve corresponds to the plasma in a pure Ar atmosphere. The dashed lines indicate the isopower curves (see gray labels). Numbers indicate the $T_{\mathrm{c}}$ values of the films 15-nm-thick deposited in the specific conditions. (b) Normalized resistance versus temperature curves for 15-nm-thick films deposited at $p_{\mathrm{Ar}}=3$~$\mu$bar, $P=$ 350 W, and varying the N$_{2}^{\%}$ content. Normalized R(T) curves of films at deposited at N$_{2}^{\%}=$ 25 both at 150 and 250 Watt are reported by thin and thick violet line, respectively.} 
\label{serieB}
\end{figure*}

NbReN films were deposited on Si substrates in a UHV DC magnetron sputtering system with a base pressure of about $2 \times 10^{-8}$ mbar. A stoichiometric NbRe (Nb$_{0.18}$Re$_{0.82}$) target (99.95\% pure, 5 cm in diameter) was reactively sputtered in a mixture of argon (Ar, 99.95\% pure, inert gas) and nitrogen (N$_2$, 99.95\% pure, reactive gas), whose relative concentration was finely tuned using two mass flow controllers. The total deposition pressure, $p_{\mathrm{tot}}$, was measured by a capacitive gauge. Finally, the sputtering was performed in a power bias mode. 

Following Refs.~\cite{Hallett2021,Bos,Henrich2012}, the analysis of the discharge parameters was performed. Indeed, different deposition conditions result in different plasma properties. In particular, the influence of the amount of nitrogen gas and the sputtering power ($P$), crucial in determining the film properties, was carefully examined. At this purpose, more than 100 films of different thickness ($d_{\mathrm{NbReN}}$) were grown. They were schematically divided into three sets based on the growth regime, as reported in Table \ref{tab:samples}. The series named A comprises the films analysed in Ref.~\cite{Cirillo2021}. They were deposited in the regime of high N$_{2}$ fraction [N$_{2}^{\%}$=N$_{2}$/(N$_{2}$+Ar)] and low $P$. On the contrary, series named as B (C) was obtained for intermediate (low) N$_{2}^{\%}$ values and high (low) $P$. After optimizing the deposition conditions for each set, we deposited films with thicknesses down to the ultrathin limit as required for the realization of SNSPDs. All depositions were performed in the regime corresponding to the formation of the nitride compound at the target~\cite{Ohring}. During this process, the target voltage ($V$) changes accordingly. In particular, once fixed both the Ar pressure ($p_{\mathrm{Ar}}$) and $P$, $V$ increases as a function of the N$_{2}$ fraction~\cite{Bos,Glowacka} due to the higher impedance of the compound with respect to the metallic target~\cite{Matsunaga}. As reported in Refs.~\cite{Bos,Glowacka}, the maximum value of $T_{\rm c}$ is observed where a steep increase appears in the $V$(N$_{2}^{\%}$) dependence. This effect was initially observed in the films of the A series, as reported in Fig. S1 (see Supplementary Material, SM). However, the films of this set have lower $T_{\mathrm{c}}$ compared to the native NbRe. For example, NbReN films 8-nm-thick deposited under optimized conditions have a $T_{\mathrm{c}}$ of about 4 K~\cite{Cirillo2021}, whereas the NbRe films of the same thickness used to realize the first SNSPDs have $T_{\mathrm{c}}=6$~K~\cite{Cirillo2020}. 

\begin{figure*}[!ht]
\centerline{\includegraphics[width=18.5cm]{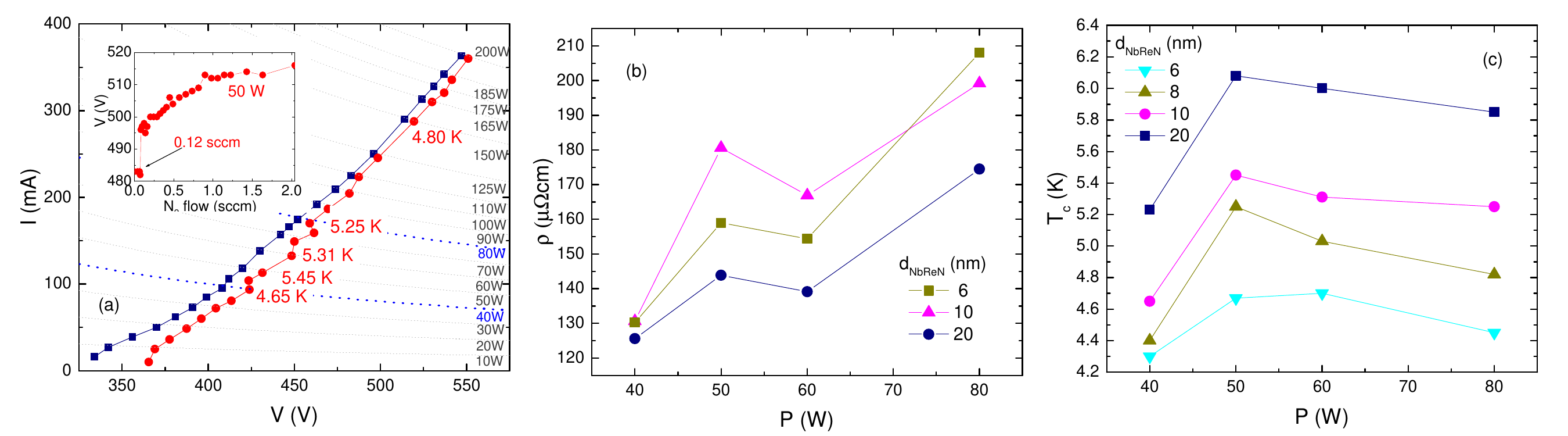}}
\caption{(a) Current-Voltage characteristics for $p_{\mathrm{Ar}}=3.5$~$\mu$bar. The navy (red) curve corresponds to the plasma in a pure Ar (for a nitrogen flow of 0.12 sccm, where the jump as a function of the nitrogen flow appears, see inset). The gray lines indicate the isopower curves (see gray labels). The blue isopower lines define the region of negative slope (see text). Numbers indicate the $T_{\mathrm{c}}$ values of the films deposited in some specific conditions for $d_{\mathrm{NbReN}}=10$~nm. (b) Dependence of low temperature resistivity on $P$ for films of different thickness, from 6 to 20~nm, deposited at a nitrogen flow in the range 0.12-0.24 sccm. (c) Dependence of $T_{\mathrm{c}}$ on the sputtering power for films of different thickness from 6 to 20~nm deposited in the same condition as panel (b).} 
\label{serieC}
\end{figure*}

In order to increase $T_c$ in the ultrathin film regime, for the new series (B and C) we analysed the discharge characteristics obtained in different sputtering regimes. The dependence of the plasma current on the discharge voltage was systematically monitored both as a function of $P$ and nitrogen content. In the case of the B series, we fixed $p_{\mathrm{Ar}}=3$~$\mu$bar. Fig.~\hyperref[serieB]{\ref*{serieB}(a)} shows the dependence of the current ($I$) on $V$, as a function of $P$ and with the addition of different N$_{2}$ fractions, from a pure Ar atmosphere (black curve, thick line) up to N$_{2}^{\%}=29$$\%$ (magenta curve). It results that $I$ is always an increasing function of $V$, with an almost linear dependence for the curve of pure NbRe, but with a small bending for the reactive processes. As discussed in Ref.~\cite{Henrich2012} in the case of NbN, the optimal deposition occurs for negative $I$-$V$ slopes, namely when a dynamic equilibrium between the nitride formation at target surface and its erosion is established. The first process is controlled by the amount of N$_{2}$ and independent on $I$, the second, on the contrary, is an increasing function of $I$ (or $P$) and dominates at high currents. In our case it was not possible to achieve the condition of the compound branch closing on the metallic one, as reported in ~\cite{Bos,Henrich2012}, probably due to the characteristics of our deposition system (target dimensions, pumping speed, and maximum $P$). Our sputtering conditions are instead more similar to what recently reported for MoN films~\cite{Hallett2021}. It is also worth noting that no hysteresis is observed in the $I$-$V$ curves, probably due to the small target diameter or the pumping speed set during the deposition~\cite{Bos}. The critical temperatures of the NbReN 15-nm-thick films deposited under the conditions described by the $I$-$V$ curves of Fig.~\hyperref[serieB]{\ref*{serieB}(a)} are reported in the graph. The $T_{\mathrm{c}}$ values were determined as the midpoint of the resistive transitions, $R(T)$. The $R(T)$ curves normalized to the value of the resistance at $T=20$ K (R$_{\mathrm{20K}}$) for the films deposited at 350~$W$ at different N$_{2}$ fractions are reported in Fig.~\hyperref[serieB]{\ref*{serieB}(b)}. In these conditions, we observe a decrease of $T_{\mathrm{c}}$ with the increase of N$_{2}^{\%}$. The values of $\rho$ of these films varies between 140 and 180 $\mu\Omega$ cm. With the nitrogen content fixed at N$_{2}^{\%}=$ 25 the dependence of $T_{\mathrm{c}}$ on $P$ was investigated at 150, 250, and 350 $W$ (see open triangles in Fig.~\hyperref[serieB]{\ref*{serieB}(a)} and violet curves in Fig.~\hyperref[serieB]{\ref*{serieB}(b)}). We notice a larger $T_{\mathrm{c}}$ value for $P=250$~$W$ (thick violet line in Fig.~\hyperref[serieB]{\ref*{serieB}(b)}), which lies in the more rounded region of the $I$-$V$ characteristic. This is in accordance with Refs.~\cite{Henrich2012}, and may be indicative of the occurrence of the dynamic equilibrium previously discussed at this $P$ value. 

\begin{figure}[!ht]
\centerline{\includegraphics[width=9cm]{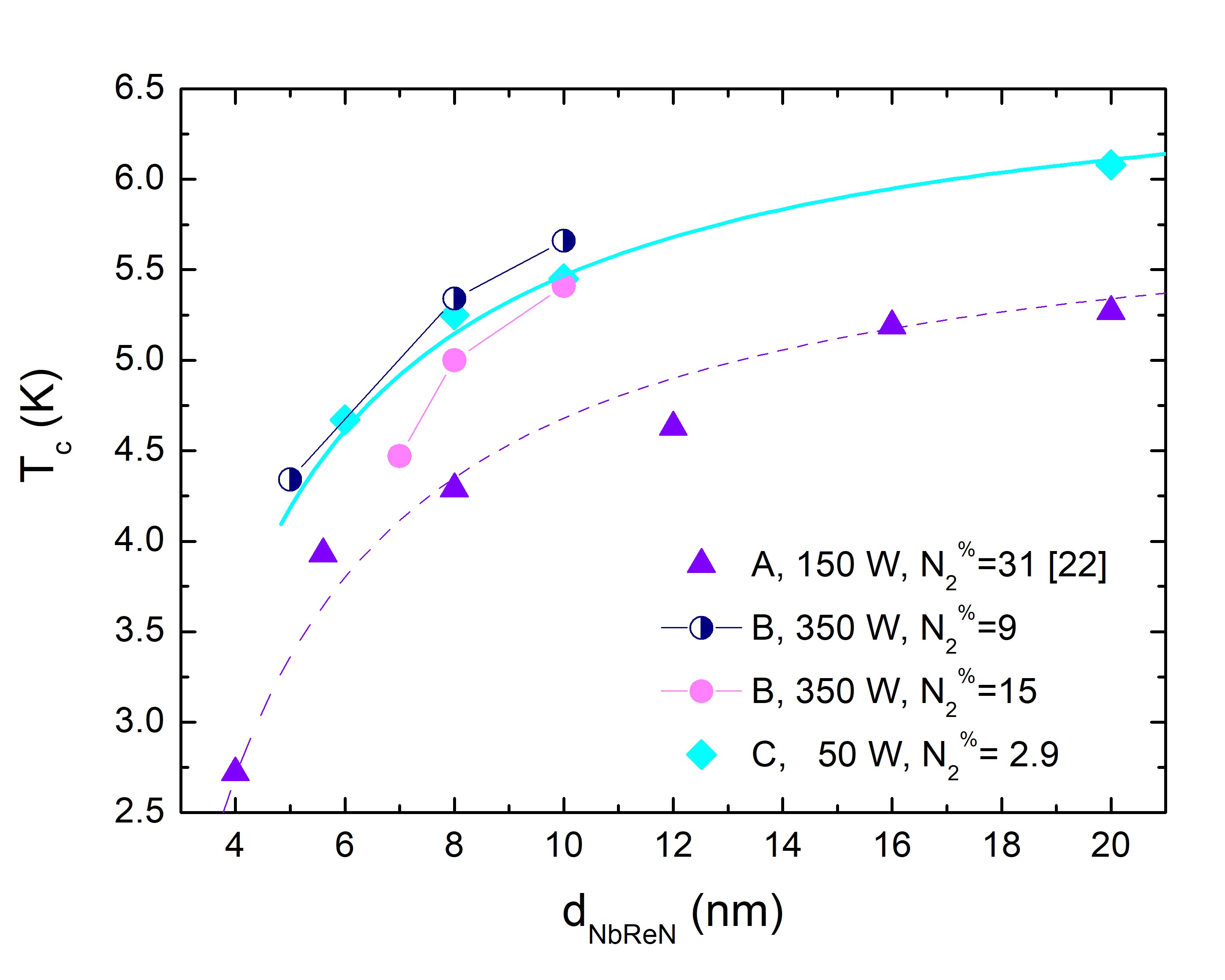}}
\caption{$T_{\mathrm{c}}(d_{\mathrm{NbReN}})$ dependence in the low thickness limit for a selection of films representative of the sample sets under study as indicated in the legend. The lines are guide to the eye.} 
\label{Tc}
\end{figure}

\begin{figure*}
    \centering
    {\includegraphics[width=18cm]{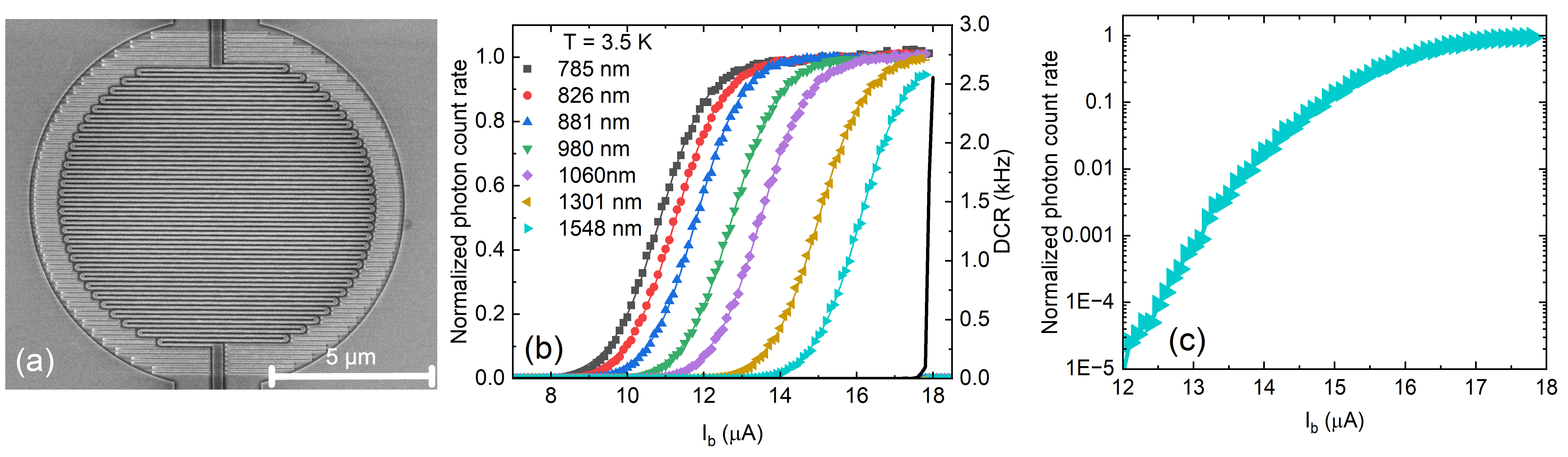}}  
    \caption{(a) SEM image of a representative SNSPD device; (b) normalized photon count rate data points (dots) and sigmoidal fittings (lines) for $\lambda$ ranging from 785~nm to 1548~nm (left axis) and DCR (right axis) at $T = 3.5$~K; (c) normalized photon count rate at $\lambda =$ 1548 nm and $T = 3.5$~K in logarithmic scale.}  
\label{D2}  
\end{figure*} 

\begin{figure}[!h] 
    \centering
    \subfloat
    {\includegraphics[width=8.4cm]{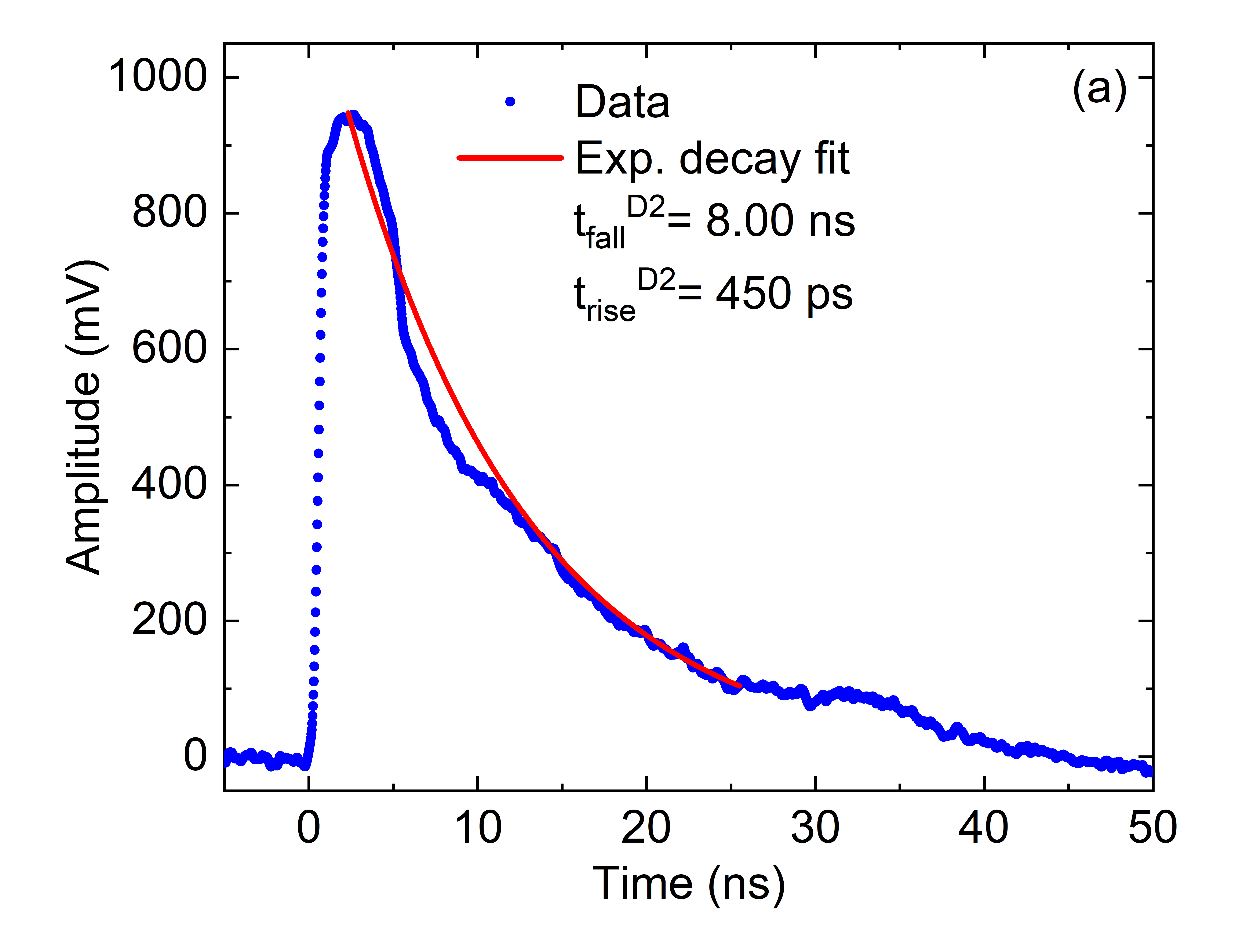}}\\
    {\includegraphics[width=9.1cm]{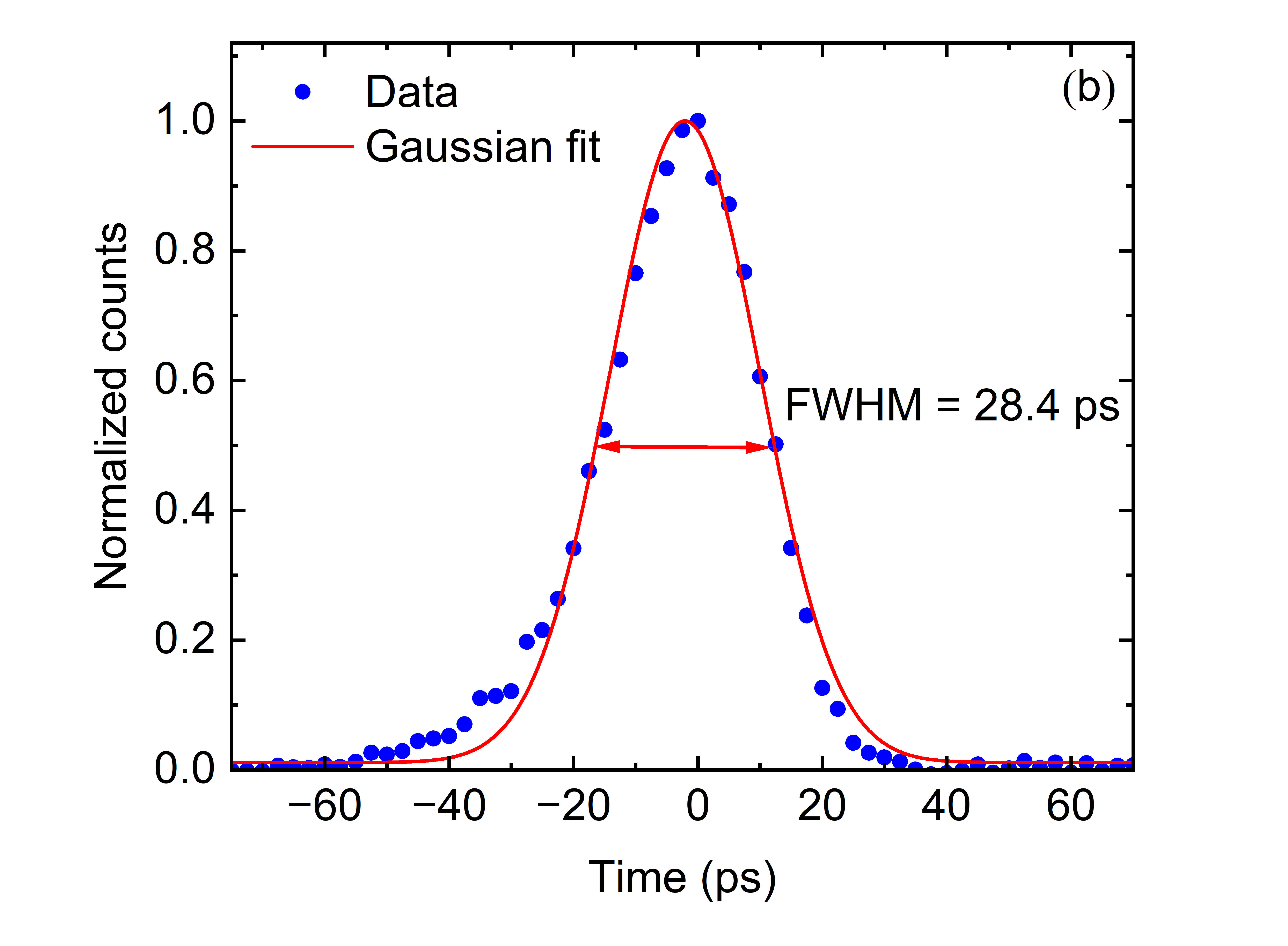}} 
    \caption{(a) Single pulse at $T = 3.5$~K (blue), with exponential fit of the decay (red); $t_{\text{rise}}$ and $t_{\text{fall}}$ are shown. (b) Timing jitter at $T = 3.5$~K for a similar detector; data (points) and Gaussian fit (line).}  
    \label{pulse}
\end{figure}

Since we could not observe the full deviation from the compound to the metallic branch in the $I$-$V$ curves in the case of series B, we decided to move to a regime of lower power and nitrogen flow (set C). We started by analyzing the $V$(N$_{2}^{\%}$) curves at different $P$. As an example, in the inset of Fig.~\hyperref[serieC]{\ref*{serieC}(a)} we plot the dependence of the plasma voltage as a function of $N_{2}^{\%}$ for $p_{\mathrm{Ar}}=3.5$~$\mu$bar and $P=50$~W. A clear jump is obtained at a flow of 0.12 sccm. A similar behavior was observed in a large range of power (data not reported here). Therefore, we restricted ourselves to this nitrogen fraction and constructed the corresponding $I$-$V$ characteristic by changing $P$. This result is shown in the main panel of Fig.~\hyperref[serieC]{\ref*{serieC}(a)}. Here, the $I$-$V$ curve in pure Ar (squares) is compared to the reactive characteristics (circles). Even if the two branches are now very close, it is possible to observe that they have different shapes. At very low power they exhibit different concavities, while they tend to merge in the high power limit. In the intermediate regime, between about 40 and 80~$W$ (as indicated by the blue isopower curves), the characteristics are more distant and a hint of a negative resistance is present. As in the case of the B series, this plot shows the values of $T_{\mathrm{c}}$ for the films with $d_{\mathrm{NbReN}}=10$~nm deposited at different powers. In agreement with the literature~\cite{Henrich2012}, larger values of the critical temperature are obtained for films grown in the regime of negative plasma resistance, namely in the region enclosed within the blue curves of Fig.~\hyperref[serieC]{\ref*{serieC}(a)}. Therefore, films of different thicknesses, from $d_{\mathrm{NbReN}}=20$~nm down to the ultrathin limit $d_{\mathrm{NbReN}}=6$~nm, were deposited in these growth conditions. We first analyzed the dependence of $\rho$ on the sputtering power, as reported in Fig.~\hyperref[serieC]{\ref*{serieC}(b)} for $d_{\mathrm{NbReN}}=6, 10$ and $20$~nm. The values of $\rho$ are comparable with those of series B, with a tendency towards increased resistivity for higher power and small $d_{\mathrm{NbReN}}$. Concerning the critical temperature, in Fig.~\hyperref[serieC]{\ref*{serieC}(c)} the $T_{\mathrm{c}}(W)$ dependence is reported for $d_{\mathrm{NbReN}}=6, 8, 10$ and $20$~nm. The curves scale without any crossings, clearly demonstrating a systematic enhancement of the critical temperature with increasing $d_{\mathrm{NbReN}}$. The behaviors are non-monotonous, since the largest $T_{\mathrm{c}}$ is obtained for the film deposited at 50~$W$ for all $d_{\mathrm{NbReN}}$. It is important to stress that by tuning the growth conditions, the values of $T_{\mathrm{c}}$ increased considerably. This is clearly evident from Fig. \ref{Tc}, where we compared the results obtained for the different series of films grown according to the more promising conditions. For example, for $d_{\mathrm{NbReN}}=8$~nm at $W=50$~$W$, it is $T_{\mathrm{c}}=5.25$~K, more than one Kelvin larger than the one of series A~\cite{Cirillo2021}. The $T_{\mathrm{c}}(d_{\mathrm{NbReN}})$ dependecies also indicate that growing the films at low $N_2^{\%}$ is particularly beneficial in the ultrathin limit, $d_{\mathrm{NbReN}} < 10$ nm, which is relevant for application as SNSPDs.

Superconducting nanowire single-photon detectors were fabricated from a 10-nm-thick NbReN film deposited under regime C at $p_{\mathrm{Ar}} = 3.5$~$\mu$bar, N$_2^{\%} =$ 2.8, and $P = 150$~W. The film shows $T_{\mathrm{c}} = 5.00$~K and resistivity $\rho = 125$~$\mu\Omega$cm. A 100~kV electron beam lithography system (Raith EBPG-5200) was used to define patterns on a 100-nm-thick AR-P 6200 resist. Subsequent structures were etched using reactive ion etching with a gas mixture of {SF$_6$}/{O$_2$} (13.5/3.5 sccm flow rates). After fabrication, the remaining resist was removed using wet cleaning (Anisole) and then SNSPDs were covered by a 12~nm layer of SiN deposited via plasma-enhanced chemical vapor deposition, to prevent oxidation. The completed devices were mounted in a Gifford-McMahon closed cycle cryostat with base temperature of 3.5 K, and characterized under flood illumination configuration. Among the geometries we tested, the devices with linewidth $70$~nm, pitch $140$~nm, and radius $4.5$~$\mu$m showed the best performances, exhibiting IDE saturation at longer wavelengths together with superior timing jitter. For this reason, the corresponding SNSPDs were chosen as representative samples. Scanning Electron Microscope (SEM) image of the selected device is shown in Fig.~\hyperref[D2]{\ref*{D2}(a)}. The critical current density of the detector at $T = 3.5$~K is $2.46\times10^{10}$~A/m$^2$. Photon count rates and dark count rates of the representative SNSPD measured as a function of bias current ($I_b$) at $T = 3.5$~K ($T/T_{\mathrm{c}}=0.7$) for wavelengths ($\lambda$) from 785~nm to 1548~nm are presented in Fig.\hyperref[D2]{\ref*{D2}(b)}. The results were normalized with respect to the saturation values of the sigmoid fitting function, and they clearly show saturated IDE for wavelengths up to 1301~nm, and 95\% IDE at 1548~nm. This represents a significant improvement with respect to the efficiency of the first NbRe-based SNSPDs. In that case, the best performing device (thickness 8 nm, linewidth $50$~nm, pitch $100$~nm, and radius $5$~$\mu$m) at the same wavelength reached a lower saturation of about 89\%, despite operating at $T/T_{\mathrm{c}}=0.5$~\cite{Cirillo2020}. In addition to the linear-scale plot, the IDE at 1550~nm is also reported in Fig. \hyperref[D2]{\ref*{D2}(c)} in logarithmic scale for a visual confirmation of the curve bending as saturation is approached. Characteristic times were measured at $I_{\text{b}} \approx 95\%\, I_{\text{c}}$ by using 1301~nm laser illumination at $T = 3.5$~K via a cryogenic amplification stage and 4~GHz bandwidth oscilloscope. Fig.~\hyperref[pulse]{\ref*{pulse}(a)} shows a single pulse with an exponential decay fit; $t_{\text{rise}} = 450$~ps, defined as the time interval between 20\% and 80\% of the pulse amplitude, and $t_{\text{fall}} = 8.00$~ns, from exponential decay fitting. The irregularities observed in the falling edge of the pulse are attributed to reflections within the readout circuitry. It is worth noting that the intrinsic fall time could be shorter than the 8 ns extracted from the exponential fit, as the initial decay appears faster and may be partially masked by reflections. The pulse shape at $T = 2.4$~K measured via a room-temperature amplification stage is reported in the SM as Fig. S3. The timing jitter of another device with the same geometry on the same chip was measured at $T = 3.5$~K, and it is reported in Fig.~\hyperref[pulse]{\ref*{pulse}(b)}. The slightly asymmetric distribution may stem from resistivity variations due to partial oxidation~\cite{Zadeh2020}. Measured jitter of this SNSPD is $28.4 \pm 0.5$~ps at full-width half-maximum (FWHM) of the Gaussian fit. These values are comparable to those of NbRe- and NbTiN-based SNSPDs at $T = 4.3$~K~\cite{Zadeh2017,Gourgues2019,Cirillo2020} and outperform some amorphous high-performance materials at a similar temperatures~\cite{Marsili2013,Verma2014_022602,Korneeva2014,Chiles2020}. Notably, these promising results can be further improved through film tuning and process optimization~\cite{Holzman2019,Zadeh2017}. \\ \indent In conclusion, NbReN ultrathin films with improved $T_{\mathrm{c}}$ with respect with the first reported samples~\cite{Cirillo2021}, were deposited by tuning the sputtering conditions. SNSPDs fabricated from these films exhibit 100\% IDE up to 1301~nm and 95\% IDE at 1548~nm, low timing jitter, and nanosecond recovery times. These results position NbReN as a promising candidate for detectors operating at cryogenic temperatures achievable with compact closed-cycle systems and potentially extendable to the MIR. Further optimization of the growth and patterning processes may lead to improved performances, broadening the applicability of this material to quantum technologies.

\section*{Supplementary Material}
See the Supplemental Materials for details on the deposition conditions of the films of the A series, resistive transitions of films belonging to the C series, and SNSPD pulse shape at $T = 2.4$ K with room temperature
amplification stage.

\begin{acknowledgments}
This research was partially supported by the QUANCOM Project (MUR PON Ricerca e Innovazione No. 2014$–$2020 ARS01$\_$00734) and the project IR0000003$-$IRIS supported by the NextGeneration EU funded Italian National Recovery and Resilience Plan with the Decree of the Ministry of University and Research number 124 (21$/$06$/$2022) for the Mission 4 $-$ Component 2 $-$ Investment 3.1.
\end{acknowledgments}

\section*{Data Availability Statement}
The data that support the findings of this study are available from the corresponding author upon reasonable request.

\end{document}